\documentclass[useAMS,usenatbib,usegraphicx]{mn2e}

\newcommand{\etalb}{et al.}
\newcommand{\beq}{\begin{equation}}
\newcommand{\beqa}{\begin{eqnarray}}
\newcommand{\eeq}{\end{equation}}
\newcommand{\eeqa}{\end{eqnarray}}
\newcommand{\del}{\delta}

\title[The Difference PDF of 21-cm Fluctuations: A Powerful Statistical
Tool for Probing Cosmic Reionization] {The Difference PDF of 21-cm
Fluctuations: A Powerful Statistical Tool for Probing Cosmic Reionization}

\author[R. Barkana and A. Loeb]{Rennan Barkana$^{1}$ and Abraham Loeb$^{2}$
\thanks{E-mail: barkana@wise.tau.ac.il (RB); aloeb@cfa.harvard.edu (AL)}\\
$^{1}$School of Physics and Astronomy, The Raymond and Beverly Sackler
Faculty of Exact Sciences,\\ Tel Aviv University, Tel Aviv 69978,
ISRAEL\\ $^{2}$Astronomy Department, Harvard University, 60 Garden
Street, Cambridge, MA 02138, USA}

\begin{document}

\pagerange{\pageref{firstpage}--\pageref{lastpage}} \pubyear{2007}

\maketitle

\label{firstpage}

\begin{abstract}
A new generation of radio telescopes are currently being built with the
goal of tracing the cosmic distribution of atomic hydrogen at
redshifts 6-15 through its 21-cm line. The observations will probe the
large-scale brightness fluctuations sourced by ionization fluctuations
during cosmic reionization. Since detailed maps will be difficult to
extract due to noise and foreground emission, efforts have focused on a
statistical detection of the 21-cm fluctuations. During cosmic
reionization, these fluctuations are highly non-Gaussian and thus more
information can be extracted than just the one-dimensional function that is
usually considered, i.e., the correlation function. We calculate a
two-dimensional function that if measured observationally would allow a
more thorough investigation of the properties of the underlying ionizing
sources. This function is the probability distribution function (PDF) of
the difference in the 21-cm brightness temperature between two points, as a
function of the separation between the points. While the standard
correlation function is determined by a complicated mixture of
contributions from density and ionization fluctuations, we show that the
difference PDF holds the key to separately measuring the statistical
properties of the ionized regions.
\end{abstract}


\section{Introduction}

The earliest generations of stars are thought to have transformed the
universe from darkness to light and to have reionized and heated the
intergalactic medium. Knowing how the reionization process happened is
a primary goal of cosmologists, because this would tell us when the
early stars formed and in what kinds of galaxies. The clustering of
these galaxies is particularly interesting since it is driven by
large-scale density fluctuations in the dark matter
\citep{BLflucts}. While the distribution of neutral hydrogen during
reionization can in principle be measured from maps of 21-cm emission
by neutral hydrogen, upcoming experiments such as the Mileura
Widefield Array\footnote{http://www.haystack.mit.edu/ast/arrays/mwa/}
and the Low Frequency Array\footnote{http://www.lofar.org/} are
expected to be able to detect ionization fluctuations only
statistically \citep[for reviews see, e.g.,][]{fob06,bl07}.

Studies of statistics of the 21-cm fluctuations have focused on the
two-point correlation function (or power spectrum) of the 21-cm
brightness temperature. This is true both for analytical and numerical
studies and analyses of the expected sensitivity of the new
experiments \citep{miguel,CfA}. The power spectrum is the natural
statistic at very high redshifts, as it contains all the available
statistical information as long as Gaussian primordial density
fluctuations drive the 21-cm fluctuations. However, during
reionization the hydrogen distribution is a highly non-linear function
of the distribution of the underlying ionizing sources. This follows
most simply from the fact that the H~I fraction is constrained to vary
between 0 and 1, and this range is fully covered in any scenario
driven by stars, in which the intergalactic medium is sharply divided
between H~I and H~II regions.  The resulting non-Gaussianity
\citep{Bha} raises the possibility of using complementary statistics to
measuring additional information that is not directly derivable from
the power spectrum \citep{Ali}.

Numerical simulations have recently begun to reach the large scales
(of order 100 Mpc) needed to capture the evolution of the IGM during
reionization \citep{mellema,zahn}. These simulations account
accurately for gravitational evolution on a wide range of scales but
still crudely for gas dynamics, star formation, and the radiative
transfer of ionizing photons. Analytically, \citet{fzh04} used the
statistics of a random walk with a linear barrier to model the H~II
bubble size distribution during the reionization epoch.  Schematic
approximations were developed for the two-point correlation function
\citep{fzh04, m05}, but recently \citet{b07} developed an accurate,
self-consistent analytical expression for the full two-point
distribution within the \citet{fzh04} model, and in particular used it
to calculate the 21-cm correlation function.

Noting the expected non-Gaussianity and the importance of additional
statistics, \citet{fzh04} also calculated the one-point probability
distribution function (PDF) of the 21-cm brightness temperature $T_b$
at a point. The PDF has begun to be explored in numerical simulations
as well \citep{ciardi,mellema}. Some of the additional information
available in the PDF can be captured by the skewness \citep{wyithe} or
bispectrum \citep{Ali} statistics. Both the correlation function and
the PDF are functions of a single variable (at each redshift): the
two-point correlation function is a function of separation, and the
PDF is a function of $T_b$. It is possible to create a two-dimensional
function by calculating the one-point PDF as a function of smoothing
scale (or pixel size), but this quantity is difficult to interpret
since it is not simply related to the 21-cm correlation function or to
the ionization statistics.

In this paper we consider a two-dimensional function that generalizes
both the one-point PDF and the correlation function and yields
additional information beyond those statistics. In particular, the
variance of this new statistic is simply related to the 21-cm
correlation function that is usually considered. This function is the
PDF of the difference $\Delta T_b \equiv T_{b,1} - T_{b,2}$ of the
21-cm brightness temperatures $T_b$ at two points. We present in the
next section our analytical model for predicting the difference PDF;
its precise relation to the two-point correlation function is
presented in section~\ref{sec:pdf}. We present illustrative
predictions of the difference PDF in section~3, where in section~3.2
we emphasize that it can be used to separately measure ionization
correlations. We summarize our conclusions in section~4.

\section{Model}

Analytical approaches to galaxy formation and reionization are based
on the mathematical problem of random walks with barriers. The
statistics of a single random walk can be used to calculate various
one-point distributions; in particular, the statistics of a random
walk with a linear barrier can be used to calculate the distribution
of ionized bubble sizes during reionization \citep{fzh04}. However, to
calculate the correlation function and other two-point distributions
requires us to solve for the simultaneous evolution of two correlated
random walks at two different points. \citet{sb} found an approximate
but quite accurate analytical solution in the case of constant
barriers and used it to calculate the joint, bivariate mass function
of halos forming at two redshifts; \citet{st} showed that this solution
describes well the two-point correlation function of halos in
numerical simulations, particularly when expressed in Lagrangian
coordinates (i.e., in terms of the initial comoving halo
separation). \citet{b07} generalized the two-point solution to the
case of linear barriers, and applied it to calculate the correlation
function of cosmological 21-cm fluctuations during reionization. In
this section we first review the basic setup of the two-barrier
problem in the context of reionization. We then briefly summarize the
solution of \citet{b07}, except that we generalize it slightly to the
case of measurements in the presence of additional smoothing (e.g.,
due to a limited instrumental resolution).  We show how to apply this
solution to calculate the difference PDF of 21-cm fluctuations.

\subsection{Reionization: basic setup}

The basic approach for using random walks with barriers in cosmology
follows \citet{bc91}, who used it to rederive and extend the halo
formation model of \citet{ps74}. In this approach we work with the
linear overdensity field $\del({\bf x},z) \equiv \rho({\bf
x},z)/\bar\rho(z) - 1$, where ${\bf x}$ is a comoving position in
space, $z$ is the cosmological redshift and $\bar \rho$ is the mean
value of the mass density $\rho$. In the linear regime, the
overdensity grows in proportion to the linear growth factor $D(z)$
(defined relative to $z=0$). This fact is used in order to extrapolate
the linear density field to the present time, i.e., the initial
density field at high redshift is extrapolated to the present by
multiplication by the relative growth factor. We adopt this view, and
throughout this paper quantities such as $\del$ and the power spectrum
$P(k)$ refer to their values linearly-extrapolated to the present.  In
each application there is in addition a barrier that signifies the
critical value which the linearly-extrapolated $\delta$ must reach in
order to achieve some physical milestone on some scale. In this work
the milestone corresponds to having a sufficient number of galaxies
within some region in order to fully reionize that same region.

At a given $z$, we consider the smoothed density in a region around a
fixed point $A$ in space. We begin by averaging over a large scale or,
equivalently, by including only small comoving wavenumbers $k$. We
then average over smaller scales (i.e., include larger $k$) until we
find the largest scale on which the averaged overdensity is higher
than the barrier; in the application to reionization, we then assume
that the point $A$ belongs to an H~II bubble of this
size. Mathematically, if the initial density field is a Gaussian
random field and the smoothing is done using sharp $k$-space filters,
then the value of the smoothed $\del$ undergoes a random walk as the
cutoff value of $k$ is increased. Instead of using $k$, we adopt the
(linearly-extrapolated) variance $S$ of density fluctuations as the
independent variable. While the solutions are derived in reference to
sharp $k$-space smoothing, we follow the traditional extended
Press-Schechter approach and substitute real-space quantities in the
final formulas. In particular, $S$ is calculated as the variance of
the appropriate mass $M$ enclosed in a spatial sphere of comoving
radius $r$.

We apply mathematical random-walk statistics to the distribution of H
II regions during reionization using the model of \citet{fzh04}.
According to this model, a given point $A$ is contained within a
bubble of size given by the largest surrounding spherical region that
contains enough ionizing sources to fully reionize itself. If we
ignore recombinations, then the ionized fraction in a region is given
by $\zeta f_{\rm coll}$, where $f_{\rm coll}$ is the collapse fraction
(i.e., the gas fraction in galactic halos) and $\zeta$ is the overall
efficiency factor, which is the number of ionizing photons that escape
from galactic halos per hydrogen atom (or ion) contained in these
halos. This simple version of the model remains approximately valid
even with recombinations if the number of recombinations per hydrogen
atom in the IGM is roughly uniform; in this case, the resulting
reduction of the ionized fraction by a constant factor can be
incorporated into the value of $\zeta$.

In the extended Press-Schechter model \citep{bc91},
in a region containing a mass corresponding to variance $S$,
\beq f_{\rm coll} = {\rm erfc} \left(\frac{\del_c(z) - \del} {\sqrt{2 
(S_{\rm min} - S)}} \right) \ , \eeq where $S_{\rm min}$ is the
variance corresponding to the minimum mass of a halo that hosts a
galaxy, $\del$ is the mean density fluctuation in the given region,
and $\del_c(z)$ is the critical density for halo collapse at
$z$. While this describes fluctuations in $f_{\rm coll}$ well, the
cosmic mean collapse fraction (and thus the overall evolution of
reionization with redshift) is better described by the halo mass
function of \citet{shetht99} (with the updated parameters suggested by
\citet{st02}). We thus use the latter mean mass function and adjust
$f_{\rm coll}$ in different regions in proportion to the extended
Press-Schechter formula; \citet{BLflucts} suggested this hybrid
prescription and showed that it fits a broad range of simulation
results. With these assumptions, the exact ionized fraction $x^i$ in a
region is given by
\beq \bar{x}^i_{\rm exact} (z,S_{\rm min},\del, S) = \zeta\, \frac
{\bar{f}_{\rm ST}} {\bar{f}_{\rm PS}}\, {\rm erfc}
\left(\frac{\del_c(z) - \del} {\sqrt{2 (S_{\rm min} - S)}} \right) \ ,
\label{eq:xiex} \eeq where $\bar{f}_{\rm ST}$ and $\bar{f}_{\rm PS}$
are the cosmic mean collapse fractions according to the Sheth-Tormen
and Press-Schechter models, respectively.

The resulting condition $x^i=1$ for having an ionized bubble of a
given size, written as a condition for $\del$ vs.\ $S$, is of the same
form as in \citet{fzh04}, at a given redshift, and thus (as they
showed) yields a linear barrier to a good approximation (see also
\citet{fmh06}). We write the effective linear barrier in a general 
notation: \beq \delta = \nu + \mu S\ , \eeq where 
\beqa \nu & = & \del_c(z) - \sqrt{2 S_{\rm min}}\, {\rm erfc}^{-1} 
\left(\frac {\bar{f}_{\rm PS}} {\zeta \bar{f}_{\rm ST}} \right) 
\nonumber \\ \mu & = & \frac{1} {\sqrt{2 S_{\rm min}}}\, {\rm 
erfc}^{-1} \left(\frac {\bar{f}_{\rm PS}} {\zeta \bar{f}_{\rm ST}}
\right)\ , \eeqa with erfc$^{-1}$ denoting the inverse function
of erfc. For consistency, in what follows we use a modified formula
for the ionized fraction, replacing equation~(\ref{eq:xiex}) with the
expression that corresponds to the linear approximation of the
barrier:
\beq \bar{x}^i (z,S_{\rm min},\del, S) = \zeta\, \frac{\bar{f}_{\rm ST}} 
{\bar{f}_{\rm PS}}\, {\rm erfc} \left[\frac{\del_c(z) - \del} {\sqrt{2
S_{\rm min}} \left(1 - \frac{S} {2 S_{\rm min}}\right)} \right] \ . 
\label{eq:xi} \eeq This replacement ensures that the ionized fraction
varies from 0 to 1 as $\delta$ goes from $-\infty$ up to the
barrier. We also denote the neutral fraction $x^n=1-x^i$ and, in
particular, \beq
\bar{x}^n (z,S_{\rm min},\del, S) \equiv 1 -
\bar{x}^i (z,S_{\rm min},\del, S)\ . \label{eq:xn} \eeq
The approximation of the linear barrier is quite accurate as long as
the maximum $S$ that we consider is much smaller than $S_{\rm min}$,
which is the case in the applications of the model in this paper,
where the maximum $S$ is set by the resolution of the upcoming
experiments (see the next subsection).

Because of some approximations in this model, the total ionized
fraction as given by the model [see equation~(\ref{eq:meanxi}) in the
next subsection] comes out slightly different from the direct result
for the mean global ionized fraction, $\langle x^i \rangle = \zeta
\bar{f}_{\rm ST}$ in terms of the cosmic mean collapse fraction (Note
that the model of \citet{fzh04} suffers from a similar difficulty). To
deal with this, we adopt the direct values of $\langle x^i \rangle$
versus redshift, and adjust $\zeta$ within the model to an effective
value of $\zeta$ at each redshift that gives a model value of $\langle
x^i \rangle$ that equals the desired one. This typically only requires
an adjustment by a few percent or less.

\subsection{The 21-cm one-point PDF}

Before considering two-point functions, we first calculate the 21-cm
PDF around one point by following \citet{fzh04}, except that we obtain
the ionized fraction from eq.~(\ref{eq:xi}) for consistency with the
barrier, and we also apply a non-linear correction to the density. We
denote the PDF itself by $P_{T_b}$, and the cumulative probability
distribution (CPD) $\int P_{T_b} d T_b$ by $C_{T_b}$.

During cosmic reionization, we assume that there are sufficient
radiation backgrounds of X-rays and of Ly$\alpha$ photons so that the
cosmic gas has been heated to well above the cosmic microwave
background temperature and the 21-cm level occupations have come into
equilibrium with the gas temperature. In this case, the observed 21-cm
brightness temperature relative to the CMB is independent of the spin
temperature and, for our assumed cosmological parameters, is given by
\citep{Madau} \beq T_b = \tilde{T}_b \Psi;\ \ \ \ \tilde{T}_b = 25
\sqrt{\frac{1+z} {8}}\, {\rm mK}\ , \label{eq:Tb}\eeq with $\Psi = x^n
[1+\delta^{\rm L}(z)]$, where $x^n$ is the neutral hydrogen fraction and
$\delta^{\rm L}(z) \equiv D(z) \del$ is the linear overdensity at $z$ (as
opposed to $\del$ which denotes the density linearly-extrapolated to
redshift 0). Under these conditions, the 21-cm fluctuations are thus
determined by fluctuations in $\Psi$.

In the model, $x^n$ is determined by the halo abundance, which is in
turn determined by the statistics of the linear density field, and
thus $x^n$ naturally falls within the correct range of 0 to
1. However, $\Psi$ also depends on the actual density, and the linear
density $\delta^{\rm L}(z)$ can take on unphysical values below
-1. While a full non-linear model would be difficult to solve, within
the context of the model where statistics are averaged over spherical
regions, we can make a simple, approximate correction in order to get
reasonable values for the actual, nonlinear density $\delta^{\rm
NL}(z)$. \citet{mw96} developed such an approximate formula for
$\delta^{\rm L}$ as a function of $\delta^{\rm NL}$, based on
spherical collapse (of overdensities) or spherical expansion (of
voids). In particular, they incorporated the asymptotic limits of
$\delta^{\rm NL} \rightarrow -1$ and $\delta^{\rm NL} \rightarrow
\infty$ as well as the correct behavior near $\delta^{\rm NL}=0$.
Their formula is valid for the Einstein-de Sitter universe or more
generally at high redshift (when the dynamical effect of the
cosmological constant is negligible). We similarly develop and use
here an accurate approximation for the inverse function, \beqa 1 +
\delta^{\rm NL}(\delta^{\rm L}) & = & \left( \frac{\del_c^ {\rm E} -
\delta^{\rm L}} {1.12431} \right)^{-2} + \left( \frac{\del_c^{\rm E} -
\delta^{\rm L}} {1.35} \right)^{-3/2} \nonumber \\ & & -\, 0.395223
\left( \del_c^{\rm E} - \delta^{\rm L} \right)^ {-1.72256}\ ,
\label{eq:dNL} \eeqa where $\del_c^{\rm E}=1.68647$ is the critical
collapse overdensity in an Einstein-de Sitter cosmology. Thus, the
expression we use for $\Psi$ is \beq \Psi = x^n \left \{1+\delta^{\rm
NL}[D(z) \del] \right \}\ . \label{eq:Psi} \eeq

If we denote by $P_{\Psi}$ the one-point PDF of $\Psi$ at a point,
then this is related to the 21-cm one-point PDF defined above
by \beq P_{T_b} (T_b)= \frac{1}{\tilde{T}_b} P_{\Psi}
\left( \frac{T_b} {\tilde{T}_b} \right)\ . \eeq Also, the
corresponding cumulative probability distributions are equal. We also
assume that the PDF is considered on a resolution scale $r_{\rm res}$
(corresponding to a variance $S_{\rm res}$), i.e., that the density
and ionization states are averaged on the scale $r_{\rm res}$ around
the point being considered. In other words, we are really considering
the density and ionization distributions in a region of size $r_{\rm
res}$ centered at a point.

We must now consider the separate contributions to the PDF of $\Psi$
from two possible cases. First, if the point lies within a fully
ionized region, then $\Psi$ is identically zero, so this case
contributes a $\delta_D$-function (Dirac delta function) at 0,
containing the total probability that the region is ionized. This
probability is given by the quantity $F_{\rm >,lin}(\nu,\mu,S_{\rm
res})$ in equation~(15) in \citet{b07}, as first derived by
\citet{m05}. The second case is if the point lies within a region that
is still partially neutral. In this case, $\Psi$ is given by
equation~(\ref{eq:Psi}), where for each value of $\del$ we use $x^n =
\bar{x}^n (z,S_{\rm min}, \del, S_{\rm res})$ from
equation~(\ref{eq:xn}). We then distribute the total probability of
this case into various values of $\Psi$ using the conservation of
probability, i.e., for each possible value of $\del$, the probability
$Q_{\rm lin}(\nu,\mu,\del,S_{\rm res})\, d \del$ [equation~(14) in
\citet{b07}, again in agreement with \citet{m05}] contributes to
$P_{\Psi}(\Psi)\, d \Psi$ at the value of $\Psi$ that corresponds to
$\del$. As noted by \citet{fzh04}, there is a maximum value of $\Psi$,
$\Psi_{\rm max}$, since $\Psi \rightarrow 0$ in the two extreme
limits, both when the density goes to zero (due to the density term in
$\Psi$) and when it goes up to the barrier (due to full
ionization). Thus, two values of $\delta$ contribute to each value of
$\Psi$, and $P_{\Psi}(\Psi)$ is singular at $\Psi_{\rm max}$.

\subsection{The 21-cm difference PDF}

\label{sec:pdf}

We now consider two separate points, with the same assumptions as in
the previous subsection. We wish to consider the PDF of the difference
$\Delta T_b \equiv T_{b,1} - T_{b,2}$ of the 21-cm brightness
temperatures $T_b$ at two points (or, in fact, averaged over regions
centered at each of the two points). We denote the PDF itself by
$P_{\Delta T_b}$, and the CPD $\int P_{\Delta T_b} d \Delta T_b$ by
$C_{\Delta T_b}$. If we denote by $P_{\Delta \Psi}$ the PDF of the
difference $\Delta \Psi \equiv |\Psi_1 - \Psi_2|$ between the values
of $\Psi$ in the two regions, then this is related to the 21-cm
difference PDF by \beq P_{\Delta T_b} (\Delta T_b)=
\frac{1}{\tilde{T}_b} P_{\Delta \Psi} \left( \frac{\Delta T_b}
{\tilde{T}_b} \right)\ . \eeq Also, the corresponding cumulative
probability distributions are equal.

We therefore wish to determine the PDF of the difference $\Delta \Psi$
as a function of the comoving distance $d$ between two points being
considered at redshift $z$. As before, we also assume that the PDF is
considered on a resolution scale $r_{\rm res}$, i.e., we are really
considering the joint density and ionization distributions of two
regions of size $r_{\rm res}$ centered at two points separated by a
distance $d$. The model of \citet{b07} provides the probability that
either or both of these regions lie completely within H~II bubbles
and, when the regions are not fully ionized, the model provides the
correlated distributions of their average overdensities.

We must now consider the separate contributions to the PDF of $\Delta
\Psi$ from three possible cases. First, if both points are within
fully ionized regions, then $\Delta \Psi$ is identically zero, so this
case contributes a $\delta_D$-function at 0, containing the total
probability that both regions are ionized. This probability is given
by the quantity $F_>(\nu,\nu,\mu,\mu,S_{\rm res},S_{\rm res},\xi_{\rm
res}(d))$ in equation~(40) in \citet{b07}, where $\xi$ is the
effective real-space cross-correlation between the densities of the
two regions [see section 4.1 of \citet{b07}]: \beq \xi_{\rm res}(d) =
\frac{1}{2 \pi^2} \int_0^\infty k^2 dk\, \frac{\sin(k d)} {k d} P(k)
W^2(k r_{\rm res})\ , \eeq where $W(x)$ is the Fourier transform of a
spherical top-hat window function. The second case is where one of the
regions is fully ionized, so, e.g., we assume that region 2 is in an
H~II bubble while region 1 is not, and then double the contribution in
order to include the symmetric, opposite situation. In this case,
$\Delta \Psi = \Psi_1 = x^n_1 \times \left \{ 1+ \delta^{\rm NL}[D(z)
\del_1] \right \}$, where for each value of $\del_1$ we use $x^n_1 =
\bar{x}^n (z,S_{\rm min}, \del_1, S_{\rm res})$ from
equation~(\ref{eq:xn}). We then distribute the total probability of
this case into various values of $\Delta \Psi$ using the conservation
of probability, i.e., for each possible value of $\del_1$, the
probability $f_{[\del | <>]} (\nu,\nu,\mu,\mu,\del_1,S_{\rm
res},S_{\rm res},\xi_{\rm res}(d))\, d \del_1$ [equation~(43) in
\citet{b07}] contributes to $P_{\Delta \Psi}(\Delta \Psi)\, d \Delta
\Psi$ at the value of $\Delta \Psi$ that corresponds to $\del_1$. The
final case is where both regions are not fully ionized. In this case,
$\Delta \Psi = |\Psi_1 - \Psi_2|$ is a function of $\del_1$ and
$\del_2$, and the conservation of probability turns
$Q(\nu,\nu,\mu,\mu,\del_1,\del_2,S_{\rm res},S_{\rm res},\xi_{\rm
res}(d))\, d \del_1 \, d \del_2$ [equation~(36) in \citet{b07}] into
$P_{\Delta \Psi}(\Delta \Psi)\, d \Delta \Psi$ at the appropriate
value of $\Delta \Psi$.

The variance of the 21-cm PDF (for two points separated by a distance
$d$) is \beq \langle \Delta T_b^2 \rangle = 2 \left [ \bar{\xi}
_{\Delta T_b} (0) - \bar{\xi}_{\Delta T_b}(d) \right ] \ ,
\label{eq:var} \eeq where the ordinary 21-cm correlation function is
$\bar{\xi}_{\Delta T_b}(d) = \langle \Delta T_{b,1} \Delta T_{b,2}
\rangle - \langle \Delta T_{b,1} \rangle \langle \Delta T_{b,2}
\rangle $, and the $\bar{\xi}$ notation denotes averaging on the
resolution scale $r_{\rm res}$. The variance can be calculated from
the PDF using one of these expressions: \beqa \langle \Delta T_b^2
\rangle & = & \int \left(\Delta T_b \right)^2 P_{\Delta T_b} (\Delta
T_b)\, d \Delta T_b \nonumber \\ & = & 2 \int \Delta T_b \left[ 1 -
C_{\Delta T_b}(\Delta T_b) \right] \,d \Delta T_b\ , \eeqa where we
integrated by parts to get the second expression.

To calculate the correlation function of $T_b$ (or, equivalently, of
$\Psi$) without first calculating the PDF, we can calculate various
expectation values using the \citet{b07} solution, once again
generalized to include a resolution/smoothing length. First, the mean
ionized fraction in the model is \beqa \lefteqn{\langle x^i \rangle =
F_{\rm >,lin}(\nu,\mu,S_{\rm res})}
\label{eq:meanxi} \\ &&\ \  + \int_{\del=-\infty}^{\nu + \mu S_{\rm res}} 
Q_{\rm lin}(\nu, \mu, \del,S_{\rm res})\, \bar{x}^i (z,S_{\rm
min},\del, S_{\rm res})\, d\del\ , \nonumber \eeqa where $F_{\rm
>,lin}$ and $Q_{\rm lin}$ are given in section~3 of \citet{b07}, while
the mean $\Psi$ at a point is \beqa \lefteqn{\langle \Psi \rangle = 1
- \langle x^i \rangle + } \\ \lefteqn{\int_{\del=-\infty}^{\nu + \mu
S_{\rm res}} Q_{\rm lin}(\nu,\mu, \del,S_{\rm res})\, \bar{x}^n
(z,S_{\rm min},\del, S_{\rm res})\, \delta^{\rm NL}(D(z) \del)\,
d\del\ .} \nonumber \eeqa For two points separated by a distance $d$,
\beqa \label{eq:Psi12} \lefteqn{\langle \Psi_1 \Psi_2 \rangle =
\int_{-\infty}^{\nu+\mu S_{\rm res}} d\del_1 \int_{-\infty}^{\nu+\mu
S_{\rm res} } d\del_2} \\ && \ \ \ \ \ \ \ Q(\nu,\nu,\mu,\mu,\del_1,
\del_2,S_{\rm res}, S_{\rm res}, \xi_{\rm res}(d)) \nonumber \\ && \ \
\ \ \ \ \ \times\, \left[1+ \delta^{\rm NL}(D(z) \del_1)\right]\,
\left[1+ \delta^{\rm NL}(D(z) \del_2)\right] \nonumber \\ && \ \ \ \ \
\ \ \times\, \bar{x}^n (z,S_{\rm min},\del_1, S_{\rm res})\, \bar{x}^n
(z,S_{\rm min},\del_2, S_{\rm res})\ , \nonumber \eeqa which is a
generalization of equation~(49) of \citet{b07} and reduces to that
equation in the limit $S_{\rm res} \rightarrow S_{\rm min}$. Also, in
the limit $d \rightarrow 0$ equation~(\ref{eq:Psi12}) simplifies to
\beqa \lefteqn{\langle \Psi^2 \rangle = \int_{-\infty}^{\nu+\mu S_{\rm
res}} d\del\, Q_{\rm lin}(\nu, \mu, \del,S_{\rm res})} \\ && \ \ \ \
\times\, \left \{ \left[1+ \delta^{\rm NL}(D(z) \del)\right]\,
\bar{x}^n (z,S_{\rm min},\del, S_{\rm res}) \right \}^2\ . \nonumber
\eeqa

\section{Results}

\subsection{The full 21-cm difference PDF}

In this subsection we use our model to predict the 21-cm difference
PDF, plot it in full as a two-dimensional function of the
separation $d$ and the brightness temperature difference $\Delta T_b$
of the two points, and explore its dependence on a number of the input
parameters (including the redshift and the resolution scale). In the
following subsection we then show that important information can be
extracted using gross features of the PDF that are insensitive to its
detailed shape; in particular, this information can be used to cleanly
separate out and measure statistics of the ionization field that
otherwise would be mixed in and convolved with the density field
within the usually-considered two-point correlation
function. Throughout this section, we illustrate our predictions in a
$\Lambda$CDM universe that includes dark matter, baryons, radiation,
and a cosmological constant. We assume cosmological parameters
that match the three year WMAP data together with weak lensing
observations \citep{Spergel07}, namely $\Omega_m=0.299$,
$\Omega_\Lambda=0.701$, $\Omega_b=0.0478$, $h=0.687$, $n=0.953$ and
$\sigma_8=0.826$.

\begin{figure}
\includegraphics[width=84mm]{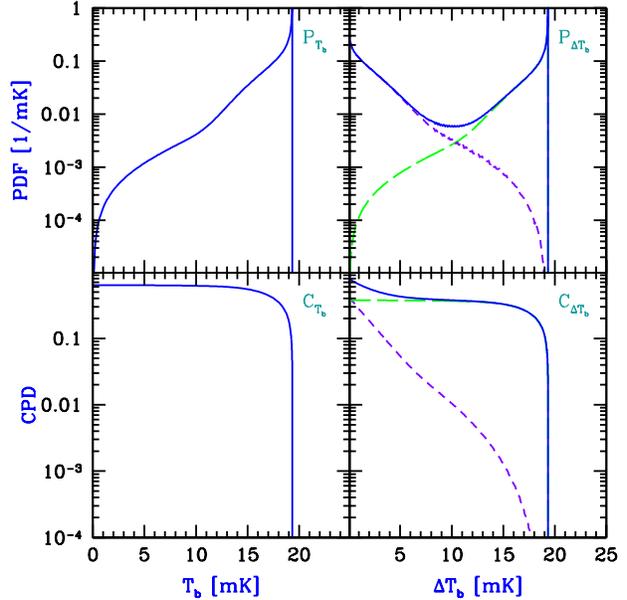}
\caption{The one-point probability distribution function (PDF) and
cumulative probability distribution (CPD) of the 21-cm brightness
temperature (panels on the left), and the two-point difference PDF and
CPD (panels on the right). We consider redshift $z=8.26$, when the
mean global ionized fraction is $\langle x^i \rangle=0.5$. In our
calculations we assume a constant ionization efficiency in all halos
in which atomic cooling occurs (i.e., in all halos above a circular
velocity $V_c = 16.5$ km/s), set so that reionization ends at
$z=6.5$. The assumed resolution (i.e., the diameter of each region) is
$1^\prime$ in all panels. Note that in each case, the PDF has an
additional $\delta_D$-function (not shown) that brings the value of
the corresponding CPD to unity at $\Delta T_b=0$ (or $T_b=0$). Also,
the PDF becomes singular at the maximum value of $\Delta T_b$ (or
$T_b$). For the two-point distributions, the full PDF or CPD (solid
curve) is the sum of contributions from both regions being neutral
(short-dashed curve) and from one of them being neutral (long-dashed
curve); the comoving separation between the two centers is set equal
to 10 Mpc.}
\label{fig:PDF}
\end{figure}

Figure~\ref{fig:PDF} shows an example of the 21-cm one-point PDF and
CPD and the two-point 21-cm difference PDF and CPD. For the one-point
function, there are two separate contributions; the case of a partly
neutral region is shown as a function of $T_b$, while the case where
the region is ionized contributes an additional $\delta_D$-function to
the PDF, or equivalently a step function to the CPD. In the CPD the
size of this step function can be easily read off as the additional
value needed to bring it up to unity at $T_b=0$. Similarly, for the
difference PDF and CPD there are three separate contributions; two of
them -- the cases of both regions being partly neutral or just one of
them -- are shown as functions of $\Delta T_b$, while the third case
-- with both regions fully ionized -- contributes an additional
$\delta_D$-function to the PDF. Again, in the CPD the size of the step
function equals the additional value needed to bring the CPD up to
unity at $\Delta T_b=0$. Another advantage of the one-point or
two-point CPDs, pointed out by \citet{fzh04} in the one-point case, is
that the PDF becomes singular (in the model) at the maximum value of
$T_b$ while the CPD does not diverge. Thus, for these two reasons, we
henceforth prefer to plot the CPD instead of the PDF.

\begin{figure}
\includegraphics[width=84mm]{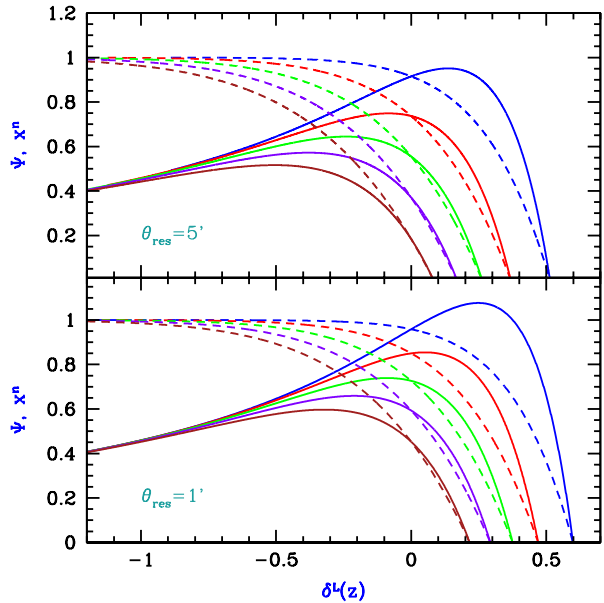}
\caption{The value of $\Psi$ and $x^n$ in a region versus the mean
linear $\delta$ in that region at redshift $z$. We show the value of
$\Psi$ as given by eq.~(\ref{eq:Psi}) [solid curves] and of the mean
neutral fraction $x^n$ in the region as given by eq.~(\ref{eq:xn})
[dashed curves]. We consider redshifts when the mean global ionized
fraction is $\langle x^i \rangle=0.1$, 0.3, 0.5, 0.7, or 0.9 (from
right to left). Our calculations assume a constant ionization
efficiency in all halos in which atomic cooling occurs, set so that
reionization ends at $z=6.5$. The assumed resolution (i.e., the
diameter of the region) is 1 or 5 arcminutes as indicated. Note that
the range where $\delta^{\rm L} < -1$ is physically sensible since it
corresponds to a $\delta^{\rm NL} > -1$.}
\label{fig:Psi}
\end{figure}

The model predicts characteristic shapes for the PDF and CPD during
reionization. In particular, the one-point function cuts off at some
maximum value of $\Delta T_b$ (which corresponds in this case to
$\Psi_{\rm max} = 0.74$), and has most of the probability near the
cutoff. The reason for this behavior is shown explicitly in
Figure~\ref{fig:Psi}. Except very early in reionization, a small (or
even somewhat negative) value of overdensity suffices in order to
significantly ionize a region. In particular, if we increase the
overdensity of a region in the model, eventually we reach a
large-enough overdensity at which the region fully reionizes itself
and $\Psi$ drops to zero. Thus it is not possible to have arbitrarily
large values of $\Psi$. In practice, $\Psi$ reaches a maximum at
$\delta^{\rm L}(z)$ values near zero. Two factors then ensure that
most of the probability of $\Psi$ values is located around this
maximum value. First, the function $\Psi$ versus $\delta^{\rm L}(z)$
is flat near its maximum (particularly in the later stages of
reionization), and second, the probability distribution of
$\delta^{\rm L}(z)$ is centered at values of $\delta^{\rm L}(z)$ near
zero. Note that we assume a Gaussian probability distribution for
$\delta^{\rm L}(z)$, although the density is weakly non-linear and its
distribution will thus be slightly modified. For instance, the
standard deviation of $\delta^{\rm L}(z)$ in these examples is $\sim
0.3$ for a $1^\prime$ resolution, and half that for a $5^\prime$
resolution. Since reionization is driven by the distribution of halos,
and the halo number density is strongly coupled to the mean density in
each region, we expect the functional form of $\Psi$ versus
$\delta^{\rm L}(z)$ to be fairly robust. This means that the shape of
the PDF will also be fairly robust even if the probability
distribution of density becomes slightly non-Gaussian.

\begin{figure}
\includegraphics[width=84mm]{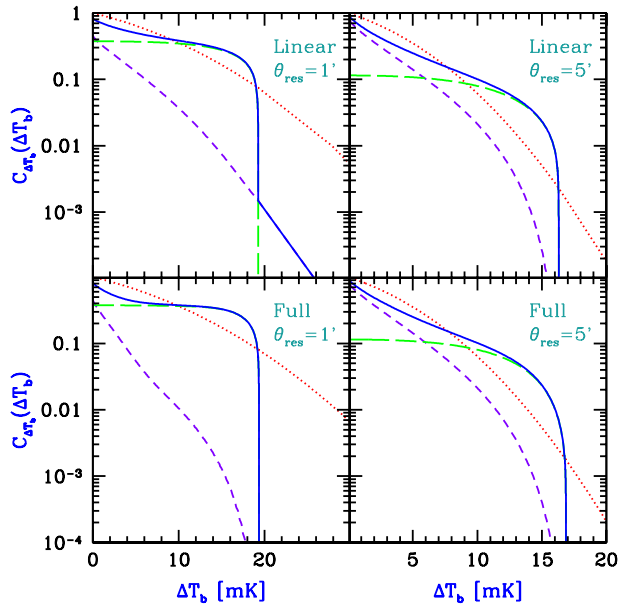}
\caption{The CPD of the 21-cm brightness temperature difference
between two spherical regions. We consider redshift $z=8.26$, when the
mean global ionized fraction is $\langle x^i \rangle=0.5$. We compare
the full CPD (solid curve), which is the sum of contributions from
both regions being neutral (short-dashed curve) and from one of them
being neutral (long-dashed curve), to the CPD of a Gaussian with the
same variance as the total CPD (dotted curve). The comoving separation
between the two centers is 10 Mpc, and the assumed resolution is
$1^\prime$ or $5^\prime$, as indicated. In each case, we show the
purely linear calculation (top panels) or the full calculation with a
non-linear correction of the density (bottom panels). Note that in
each panel, the CPD has an additional step function (not shown) that
brings its value to unity at $\Delta T_b=0$. In our calculations we
assume a constant ionization efficiency in all halos in which atomic
cooling occurs (i.e., in all halos above a circular velocity $V_c =
16.5$ km/s), set so that reionization ends at $z=6.5$.}
\label{fig:CPD1}
\end{figure}

Returning to Figure~\ref{fig:PDF}, we see that in the plotted case, the
contribution of the one-neutral-one-ionized case to the difference PDF
is similar (though not identical) in shape to the one-point PDF, and
in particular is centered near the maximum value of $\Delta T_b$. This
results from the fact that in this case, the value of $\Delta T_b$ is
simply equal to $T_b$ for the region that is not fully ionized. Note
that in general the maximum value of $\Delta T_b$ for the difference
PDF is equal to the maximum value of $T_b$ for the one-point PDF,
since $\Psi$ is always non-negative. The both-neutral contribution to
the difference PDF is quite different, since $\Psi$ in each of the two
regions tends toward $\Psi_{\rm max}$, so the most likely difference
between the two $\Psi$ values is zero. This tendency is further
strengthened by the correlation between the densities in the two
regions. The outcome of all of this is a difference PDF that has two
peaks, with a valley at intermediate values of $\Delta T_b$. As a
result, the 21-cm difference CPD first declines at small values of
$\Delta T_b$ (where it is dominated by the both-neutral case), then
flattens at larger values, and finally cuts off sharply at the maximum
value of $\Delta T_b$.

The bottom panels of Figure~\ref{fig:CPD1} again show the full CPD
(with the non-linear correction of the density), but also compare it
to the cumulative of a Gaussian (i.e., an {\it error function}) with
the same variance. The CPD shape described above is clearly very
different from the {\it error function}. As illustrated by the top
panels, the non-linear correction that we have applied to the density
is important in order to ensure that $\Psi$ is always positive and
that there is a sharp cutoff at a maximum value of $\Delta T_b$ (while
a calculation with linear densities leads to the unphysical result of
having negative values of $\Psi$ when the density fluctuation is more
negative than -1). Other than this cutoff, the non-linear correction
modifies the shape of the CPD only slightly, with the correction
having a smaller effect in the case where the resolution angle is
larger (and where fluctuations on the corresponding scale are more
linear). Thus, while a non-linear correction of the density is
required to ensure a physical result (with a non-negative density), we
do not expect our results to depend strongly on the precise form of
non-linear correction that we have used.

\begin{figure}
\includegraphics[width=84mm]{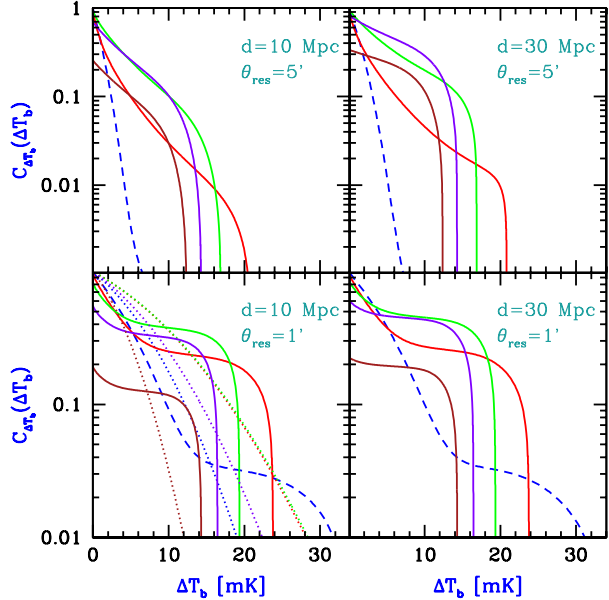}
\caption{The CPD of the 21-cm brightness temperature difference between 
two spherical regions. We consider redshifts when the mean global
ionized fraction is $\langle x^i \rangle=0.1$ (dashed curve) or
$\langle x^i \rangle=0.3$, 0.5, 0.7, or 0.9 (solid curves, from right
to left where the CPD is small). The comoving separation between the
two centers is $d=10$ Mpc or 30 Mpc, and the assumed resolution is
$1^\prime$ or $5^\prime$, as indicated. In the bottom-left panel, we
also compare the CPD curves to CPDs of Gaussians with the same
variance (dotted curves, in order from right to left: $\langle x^i
\rangle=0.5$, 0.3, 0.7, 0.1, and 0.9). Note that the CPD has an
additional step function that brings its value to unity at $\Delta
T_b=0$. In our calculations we assume a constant ionization efficiency
in all halos in which atomic cooling occurs, set so that reionization
ends at $z=6.5$.}
\label{fig:CPD2}
\end{figure}

Figure~\ref{fig:CPD2} shows the time evolution of the CPD during
reionization, considered at two different values of the separation
$d$. Throughout the parameter range considered, the CPD clearly has
the same characteristic shape as noted above (although for $\langle
x^i \rangle=0.1$ in the top panels, the flat portions occur at lower
values of the CPD than are included in the plot). In the bottom-left
panel, we show an example of two cases ($\langle x^i \rangle=0.3$ and
0.5) which have nearly identical variances (i.e., the corresponding
Gaussian CPD curves are nearly indistinguishable), and yet the actual
21-cm difference CPDs differ substantially in these two cases. The
Figure illustrates how the CPD evolves during reionization, declining
with time at low $\Delta T_b$ values (since the probability associated
with both regions being fully ionized increases), and cutting off at a
lower value of $\Delta T_b$ (since the overdensity needed for full
reionization of a given region declines as reionization progresses
globally -- see Figure~\ref{fig:Psi}).
 
\begin{figure}
\includegraphics[width=84mm]{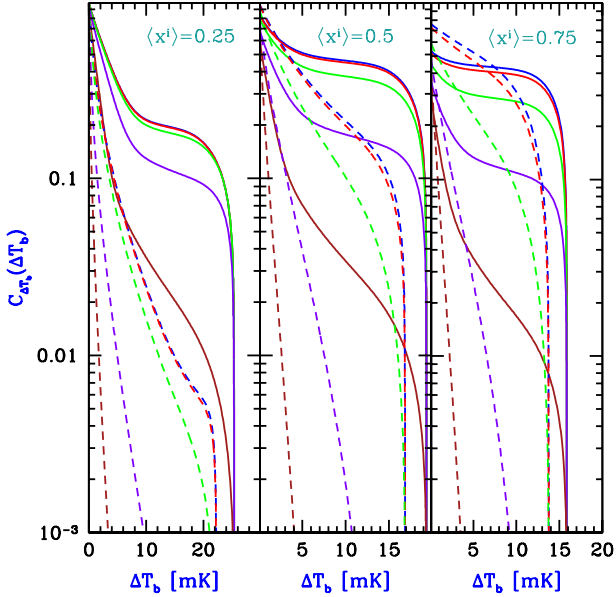}
\caption{The CPD of the 21-cm brightness temperature difference between 
two spherical regions. We consider redshifts when the mean global
ionized fraction is $\langle x^i \rangle=0.25$, 0.5, or 0.75, as
indicated. We assume a resolution of $1^\prime$ (solid curves) or
$5^\prime$ (dashed curves), at a comoving separation $d=100$, 30, 10,
3, and 1 Mpc (from right to left in each set of curves). Note that the
CPD has an additional step function that brings its value to unity at
$\Delta T_b=0$. In our calculations we assume a constant ionization
efficiency in all halos in which atomic cooling occurs, set so that
reionization ends at $z=6.5$.}
\label{fig:CPD3}
\end{figure}

The information on spatial correlations contained within the CPD is
illustrated more clearly in Figure~\ref{fig:CPD3}. In most of the
cases shown, the $d=30$ Mpc and $d=100$ Mpc curves are nearly
indistinguishable, since even at a 30 Mpc separation the two regions
are nearly independent. The CPD drops rapidly as the separation is
decreased, with the probability becoming concentrated near $\Delta
T_b=0$ once the two regions become highly correlated. The decline with
separation, which occurs at $d=1$--10 Mpc early in reionization
($\langle x^i \rangle=0.25$) but over a broader range of $d=1$--30 Mpc
later on ($\langle x^i \rangle=0.75$), indicates the relative
importance of bubble and density correlations on various scales.

\begin{figure}
\includegraphics[width=84mm]{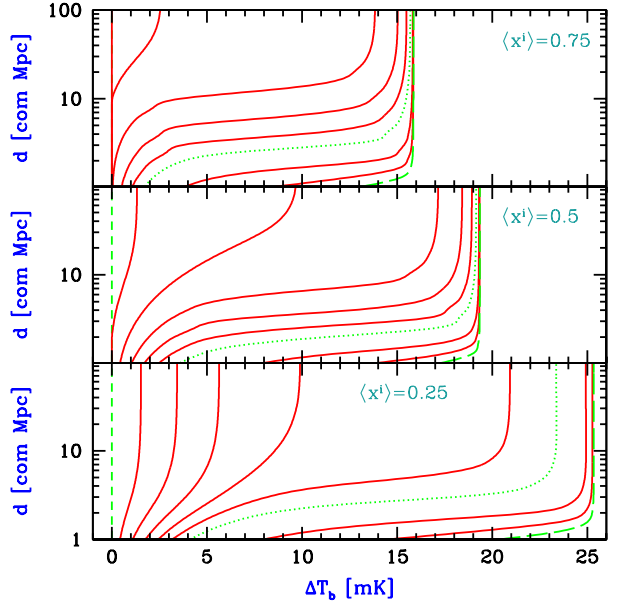}
\caption{Contour plot of the CPD of the 21-cm brightness temperature 
difference between two spherical regions. Same data as represented in
the previous Figure, but in a contour plot which emphasizes the fact
that a two-dimensional dataset, $C_{\Delta T_b}$ as a function of
$\Delta T_b$ and $d$, is available to measure at each redshift. We
again consider $\langle x^i \rangle=0.25$, 0.5, or 0.75, as
indicated. Contours go from a value $C_{\Delta T_b}=1$ (short-dashed
curve) to $C_{\Delta T_b}=0.01$ (long-dashed curve), passing through
$C_{\Delta T_b}=0.1$ (dotted curve). Equal logarithmic spacing of the
values of the CPD is used within each decade, with finer spacing at
the highest decade of CPD values (13 contours total in each panel). We
show the case of a $1^\prime$ resolution, and again assume a constant
ionization efficiency in all halos in which atomic cooling occurs, set
so that reionization ends at $z=6.5$.}
\label{fig:CPD4}
\end{figure}

A full measurement of the CPD would yield a two-dimensional function
of $d$ and $\Delta T_b$ at each redshift. This full function is
illustrated with a contour plot in Figure~\ref{fig:CPD4}. Regions that
contain much of the probability -- i.e., where the CPD changes rapidly
and there are large spaces between consecutive contours -- indicate
both the characteristic scale $d$ of correlations and a corresponding
characteristic value of $\Delta T_b$ (which is related to the
correlated distribution of densities in the two separated
regions). While the full 21-cm difference PDF would be a great tool to
study theoretically and to observe, in the next subsection we show
that even if the gross features of the PDF were measured, they already
would reveal important information that is not directly available from
measurements of the correlation function alone.

\subsection{Using the difference PDF to separately measure ionization 
statistics}

Next, we consider robust ways to extract information from the
one-point PDF and the two-point, difference PDF. While further study
may show that the detailed shape of the PDF can be used as a sensitive
probe of the underlying astrophysics (such as properties of the
population of ionizing sources), the analytical model we consider in
this paper is approximate and neglects some non-linear corrections and
other physical effects, so the precise shape would change somewhat in
more complete models or numerical simulations. However, we expect our
model to correctly capture the gross features of the PDF, which likely
constitute more robust predictions. Also, on the observational side,
while the full PDF may be difficult to measure with low
signal-to-noise data, we expect it to be easier to extract just these
gross features. We leave for future study the question of how these
features can be extracted in a realistic scenario with noise (assuming
that the PDF of the noise can first be measured accurately). Our goal
here is to study what information can be extracted just from the gross
features of the PDF.

We first consider the one-point PDF. Two gross quantities can be
simply extracted from it: the probability $P(x^i=1)$ that a region of
the resolution size is fully ionized, and the value of $\Psi_{\rm
max}$. The first quantity can be extracted by measuring the size of
the step function of the CPD, or (equivalently) by subtracting from
unity the value of the CPD just above $T_b=0$; alternatively, this
value equals the integrated area under the PDF curve, not including
the $\delta_D$-function at zero. Since most of the probability lies
near the maximum value of $T_b$ (see Figure~\ref{fig:PDF}), measuring
it depends mostly on pixels with the highest signal-to-noise
ratios. Measuring the value of $\Psi_{\rm max}$ also depends on the
same pixels; this value can be derived from the maximum value of $T_b$
using eq.~(\ref{eq:Tb}). Note that while a fully realistic PDF may not
feature such a total sudden drop at a maximum value of $T_b$, there
should nonetheless be a definite, sharp drop due to the strong
dependence of ionization on density through variations in the halo
number density. Note also that in simulations, while a region cannot
be truly fully ionized (because of the presence of very high-density
gas), the interpretation of $P(x^i=1)$ in this case is where the
bubbles within the region have fully overlapped and all the gas is
highly ionized except for some gas at $\delta \gg 1$ (which at high
redshift generally makes up only a small volume and mass
fraction). Indeed, in simulations by \citet{mellema} the PDF as a
function of $T_b$ has a fairly flat portion (though not always
increasing) followed by a rather sharp cutoff. We note that the shape
of such statistics as measured in simulations has not yet been
physically justified or been subjected to numerical convergence tests.
But it is generically expected that the two quantities $P(x^i=1)$ and
$\Psi_{\rm max}$ should clearly feature in the PDF. If models or
simulations can reliably establish at least the approximate form of
the PDF, then this would make it easier to measure these two
quantities.

Now consider the difference PDF (see Figure~\ref{fig:PDF}). One gross
feature is obviously a cutoff that can also be used to measure the
same value of $\Psi_{\rm max}$. The difference PDF can also be used to
extract three other inter-related quantities, each as a function of
the separation $d$: the probability $P(x^i_1=1,x^i_2=1)$ of joint full
ionization; the probability $2 P(x^i_1=1,x^i_2<1)$ of full ionization
of only one of the two regions (with the factor of 2 accounting for
the symmetry of selecting which region is the ionized one); and the
probability $P(x^i_1<1,x^i_2<1)$ that neither region is fully
ionized. The first probability can be extracted from the size of the
step function of the CPD at $\Delta T_b = 0$, while the other two can
be extracted from the areas under the two peaks that are fairly well
separated in the PDF during reionization (see Figure~\ref{fig:PDF}),
or more accurately by modeling the two separate contributions to the
PDF at $\Delta T_b > 0$. In reality, the three probabilities need not
be measured separately, since they can easily be shown to be closely
related to each other; in fact, $P(x^i_1=1,x^i_2=1)$ together with the
one-point quantity $P(x^i_1=1)$ can be used to express the other two
two-point probabilities: \beq P(x^i_1<1,x^i_2<1) = 1 +
P(x^i_1=1,x^i_2=1) - 2 P(x^i=1)\ , \eeq and \beq 2 P(x^i_1=1,x^i_2<1)
= 2 \left [P(x^i=1) - P(x^i_1=1,x^i_2=1) \right]\ . \eeq These
relations should make it much easier to extract this information from
even approximate measurements of the difference PDF.

Thus, from the one-point and two-point PDF, we can measure two
independent probabilities that depend directly only on ionization
statistics, not mixed in with the value of the density. These are
$P(x^i=1)$ (a single quantity at each redshift) and
$P(x^i_1=1,x^i_2=1)$ (a function of $d$ at each redshift). Even a
rough measurement of the PDFs may yield a reasonable estimate of these
gross quantities. Note also that in the limit of infinitely good
resolution (i.e., a very small resolution scale), $P(x^i=1)$ becomes
the cosmic mean ionized fraction, and $P(x^i_1=1,x^i_2=1)$ becomes the
ionization correlation function (after subtracting the square of the
mean ionized fraction). In addition, the value of $\Psi_{\rm max}$
yields an interesting piece of information on the dependence of
ionization on the density. All of these quantities are separate from
the correlation function, which yields just one function of $d$ that
is a complicated convolution of fluctuations in density and
ionization.

\begin{figure}
\includegraphics[width=84mm]{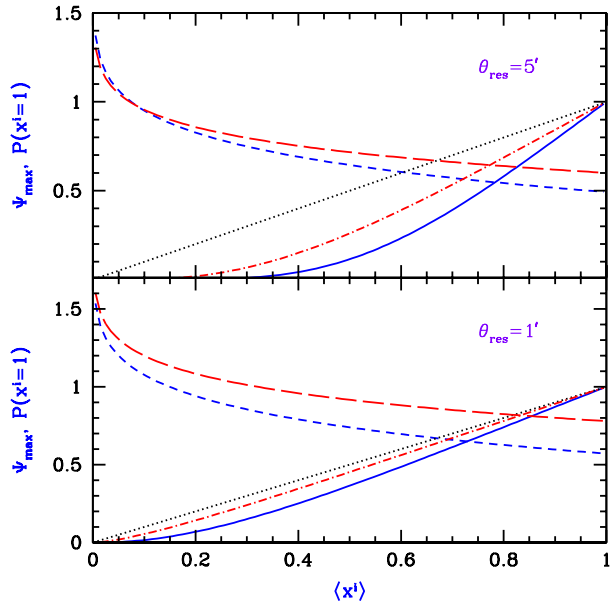}
\caption{Separate observables available from the 21-cm one-point PDF.
Assuming parameters for which the universe fully reionizes at $z=6.5$,
we consider stars forming (with a constant ionization efficiency) in
all halos above $V_c = 16.5$ km/s (corresponds to efficient atomic
cooling) or only in halos above $V_c = 45$ km/s (corresponds to strong
feedback in low-mass halos, e.g., due to photoheating or supernovae).
In each panel, we show $P(x^i=1)$ for $V_c = 16.5$ km/s (solid curve)
or $V_c = 45$ km/s (dot-dashed curve), compared to the cosmic mean
$\langle x^i \rangle$ (straight dotted line). We also show $\Psi_{\rm
max}$ for $V_c = 16.5$ km/s (short-dashed curve) or $V_c = 45$ km/s
(long-dashed curve).}
\label{fig:1pt}
\end{figure}

Figure~\ref{fig:1pt} shows predictions of our model for the two
quantities $P(x^i=1)$ and $\Psi_{\rm max}$ that are available from the
1-pt PDF. $P(x^i=1)$ can be used as a rough estimate for the cosmic
mean $\langle x^i \rangle$, although this works better with high
resolution ($1^\prime$) and late in reionization. The robustness of
theoretical predictions of the relation between $P(x^i=1)$ and
$\langle x^i \rangle$ can be investigated with further models and
simulations. Even for a fixed end-of-reionization redshift and when
expressed as functions of $\langle x^i \rangle$, both $P(x^i=1)$ and
$\Psi_{\rm max}$ depend significantly on the characteristic halo mass
of ionizing sources.

\begin{figure}
\includegraphics[width=84mm]{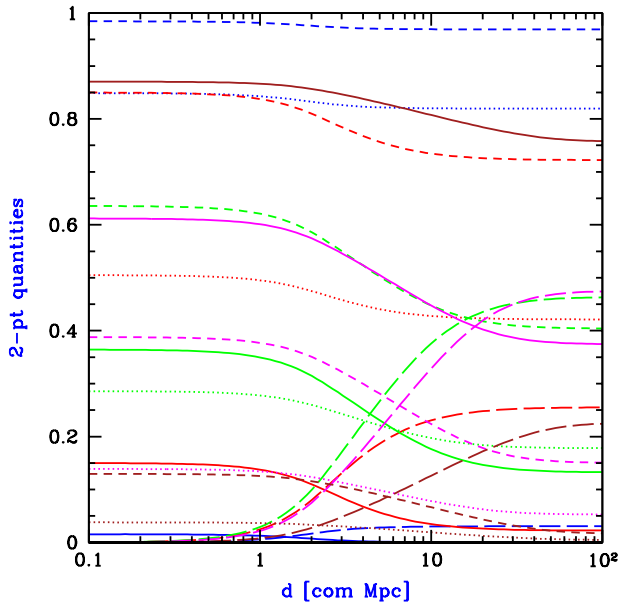}
\caption{Gross observables available from the 21-cm two-point
difference PDF, shown versus the separation of the two
points. Assuming parameters for which the universe fully reionizes at
$z=6.5$, we consider stars forming with a constant ionization
efficiency in all halos above $V_c = 16.5$ km/s. For the resolution we
assume a value of $1^\prime$. We consider at various stages during
reionization ($\langle x^i \rangle=0.1$, 0.3, 0.5, 0.7, or 0.9), the
standard quantity $\langle \Psi_1 \Psi_2 \rangle$ from the 21-cm
two-point correlation function (dotted curves, from top to bottom) as
well as the three inter-related gross quantities from the difference
PDF (which together add up to unity): $P(x^i_1=1,x^i_2=1)$ (solid
curves, from bottom to top), $2 P(x^i_1=1,x^i_2<1)$ (long-dashed
curves; at d=100 Mpc, $\langle x^i \rangle=0.1$, 0.9, 0.3, 0.5, and
0.7, from bottom to top), and $P(x^i_1<1,x^i_2<1)$ (short-dashed
curves, from top to bottom).}
\label{fig:2pt}
\end{figure}

Figure~\ref{fig:2pt} shows the three inter-related probabilities
obtainable from the difference PDF, compared to the expectation value
$\langle \Psi_1 \Psi_2 \rangle$. The standard 21-cm 2-pt correlation
function does not actually yield $\langle \Psi_1 \Psi_2 \rangle$ but
rather subtracts off $\langle \Psi \rangle^2$, which is the value of
$\langle \Psi_1 \Psi_2 \rangle$ at $d \rightarrow \infty$. Thus, in
the figures the information available from the 2-pt correlation
function is not the absolute plotted values of $\langle \Psi_1 \Psi_2
\rangle$, but just the values relative to the large-scale asymptotic
value. Therefore, the ability to measure the 2-pt correlation function
and extract useful information from it depends on the total change in
$\langle \Psi_1 \Psi_2 \rangle$ from small to large scales. This total
change is smaller than the overall change with scale of most of the
curves that show the ionization probabilities. The probability
$P(x^i_1=1,x^i_2=1)$ goes from $P(x^i_1=1)$ at $d \rightarrow 0$ to
$[P(x^i_1=1)]^2$ at $d \rightarrow \infty$, while $P(x^i_1<1,x^i_2<1)$
goes from $1-P(x^i_1=1)$ to $[1-P(x^i_1=1)]^2$. The largest change is
seen in the probability that one region is ionized and the other is
not; this varies from zero at small $d$ to $2 \{P(x^i_1=1) -
[P(x^i_1=1)]^2\}$ at large $d$.

\begin{figure}
\includegraphics[width=84mm]{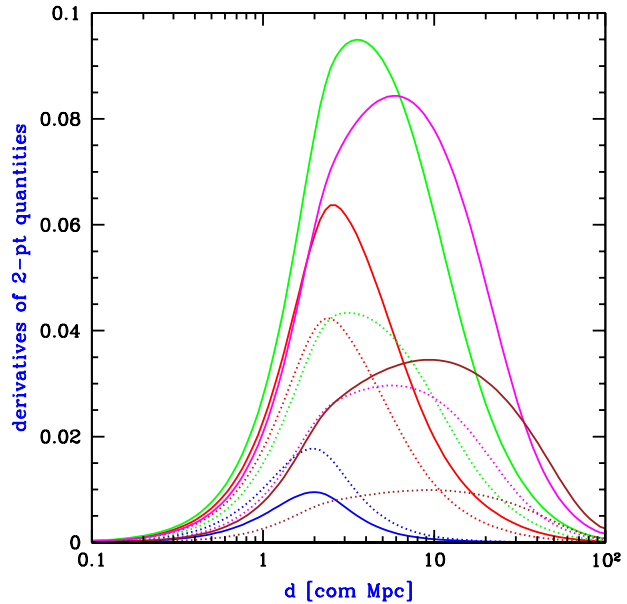}
\caption{Derivatives of the 21-cm two-point quantities. We show at
various stages during reionization $|d/d\log(d)|$ of
$P(x^i_1=1,x^i_2=1)$ (solid curves) and of $\langle \Psi_1 \Psi_2
\rangle$ (dotted curves). For each quantity, we consider various
stages during reionization ($\langle x^i \rangle=0.1$, 0.3, 0.5, 0.7,
and 0.9, in order from left to right at the peak of the curve).}
\label{fig:d2pt}
\end{figure}

\begin{figure}
\includegraphics[width=84mm]{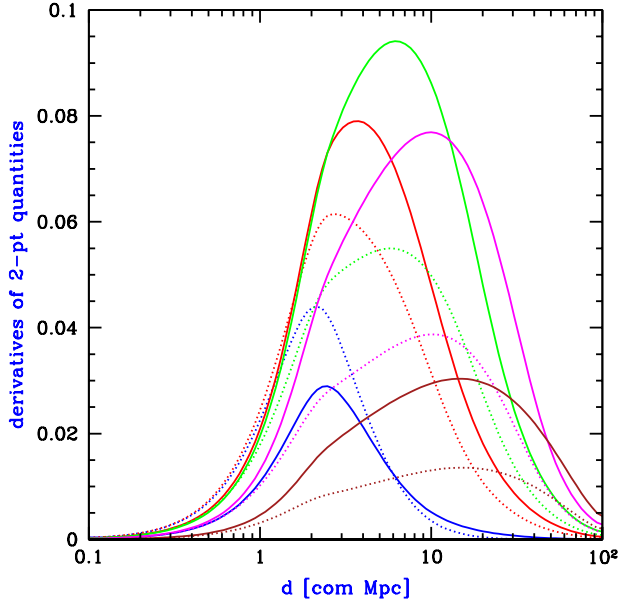}
\caption{Same as Figure~\ref{fig:d2pt}, except that we assume stars
form with a constant ionization efficiency in all halos above $V_c =
45$ km/s.}
\label{fig:d2pt45}
\end{figure}

\begin{figure}
\includegraphics[width=84mm]{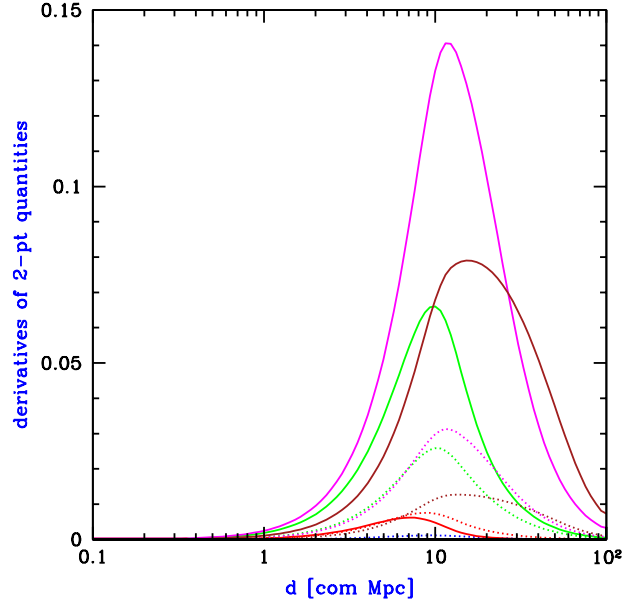}
\caption{Same as Figure~\ref{fig:d2pt}, but for the case of a
$5^\prime$ resolution. In this case, the $\langle x^i \rangle=0.1$
case cannot be seen in $P(x^i_1=1,x^i_2=1)$ (which is essentially zero
in this case, at all $d$). Here, the order of increasing peak heights
(not positions) is $\langle x^i \rangle=0.1$, 0.3, 0.9, 0.5 and 0.7,
for $\langle \Psi_1 \Psi_2 \rangle$, and $\langle x^i \rangle=0.3$,
0.5, 0.9 and 0.7 for $P(x^i_1=1,x^i_2=1)$.}
\label{fig:d2pt5}
\end{figure}

The characteristic scales of the bubble correlations can be read off
as the scales where the various quantities in Figure~\ref{fig:2pt}
change most rapidly as a function of $d$. In order to focus on this
important feature, we plot the derivatives with respect to $\log(d)$
in Figure~\ref{fig:d2pt}, just for $\langle \Psi_1 \Psi_2 \rangle$ and
for $P(x^i_1=1,x^i_2=1)$ (since the other two probabilities can be
derived from this quantity). In general, both $P(x^i_1=1,x^i_2=1)$ and
$\langle \Psi_1 \Psi_2 \rangle$ show roughly the same dominant scales,
but $P(x^i_1=1,x^i_2=1)$ shows a greater variation with scale during
the central stages of reionization at which the variation is
maximized. Figure~\ref{fig:d2pt45} shows similar trends in the case of
reionization by more massive and highly-biased halos, except that here
the characteristic scale grows to even larger values by the end of
reionization; early on in reionization, the characteristic scale is
approximately the same as in the previous case of less-massive halos,
but the magnitudes of the scale-derivatives are larger in the case of
the more-massive halos (corresponding to stronger ionization and 21-cm
fluctuations). Figure~\ref{fig:d2pt5} demonstrates the importance of
achieving high resolution in the upcoming experiments. In the scenario
considered here, a cosmic mean ionization fraction of one half occurs
at $z=8.26$, at which the resolution of $1^\prime$ corresponds to a
radius of 1.3 comoving Mpc (com Mpc), and $5^\prime$ to 6.7 com
Mpc. The typical correlation scale grows with time and overtakes the
$5^\prime$ scale only late in reionization ($\langle x^i \rangle \sim
0.7$), so until then the peaks of the curves in Figure~\ref{fig:d2pt5}
indicate roughly the resolution scale (rather than the desired
correlation scale). Observations should therefore achieve a resolution
of a few arcminutes or better in order to make it possible to measure
the evolution of the dominant correlation scale throughout most of the
reionization era.

\section{Conclusions}

We have presented and studied a new statistic for analyzing 21-cm
fluctuations, namely the PDF of the difference of the 21-cm brightness
temperatures of two regions, as a function of the separation between
their centers. This two-dimensional statistics generalizes to
one-dimensional functions, the one-point PDF and the two-point
correlation function; the latter is simply related to the variance of
the difference PDF (Eq.~\ref{eq:var}). 

We have predicted the difference PDF based on the correlated two-point
distribution of density and ionization \citep{b07}, generalized to the
case of a finite observed resolution, and including a non-linear
correction for the density (Eq.~\ref{eq:dNL}). The model predicts a
characteristic shape during reionization for the PDF (or its
cumulative form, the CPD) that can be understood from the various
contributions to it depicted in Figure~\ref{fig:PDF}. The PDF contains
information on the distribution of the density within neutral regions
and on the dominant spatial scales of bubble and density correlations
(Figures~\ref{fig:CPD2} and \ref{fig:CPD3}). The full PDF is a
function of separation and of $\Delta T_b$ at each redshift
(Figure~\ref{fig:CPD4}).

While the usual correlation function is determined by a complicated
mixture of contributions from density and ionization fluctuations, we
have shown that the difference PDF (together with the one-point PDF)
holds the key to separately measuring statistics of the ionization
distribution. In particular, even an approximately measurement of the
PDFs can generically be used to measure the ionization probability of
a resolution-sized region, and the joint ionization probability of two
such regions as a function of their separation. Within our model, the
joint ionization probability shows the same characteristic correlation
scale as the two-point 21-cm correlation function but has more power
(Figures~\ref{fig:d2pt} and \ref{fig:d2pt45}); this is because the
contributions of density and ionization to 21-cm fluctuations are
anticorrelated (a higher density implies less neutral gas), which
reduces the 21-cm fluctuations relative to ionization fluctuations.

If quasars contribute significantly to reionization by producing large
bubbles even early in reionization, or if quasars or supernovae emit
significant X-ray photons (which have a long mean free path), then the
density and ionization fluctuations will not be as simply related as
we have assumed. In this case, it will be much more important to
measure the ionization statistics separately since it will be
difficult to extract this information from the 21-cm statistics. Also
in this case the relation between the one-point and difference PDFs
and the 21-cm correlation function should be significantly different
from the case we have considered. The full quantitative details of
such a scenario goes beyond the scope of this paper and merits a
separate study.

\section*{Acknowledgments}
The authors would like to acknowledge Israel - U.S. Binational Science
Foundation grant 2004386 and Harvard University grants. RB is grateful
for the kind hospitality of the {\it Institute for Theory \&
Computation (ITC)} at the Harvard-Smithsonian CfA, where this work
began, and also acknowledges support by Israel Science Foundation
grant 629/05.

\bsp

\label{lastpage}


\begin{thebibliography}{07}

\bibitem[\protect\citeauthoryear{Barkana}{2007}]{b07} Barkana R., 
2007, MNRAS, 376, 1784

\bibitem[\protect\citeauthoryear{Barkana \& Loeb}{2004}]{BLflucts} 
Barkana R., Loeb A., 2004, ApJ, 609, 474

\bibitem[\protect\citeauthoryear{Barkana \& Loeb}{2007}]{bl07} 
Barkana R., Loeb A., 2007, Rep. Prog. Phys., 70, 627

\bibitem[\protect\citeauthoryear{Bharadwaj \& Ali}{2005}]{Bha} 
Bharadwaj, S., \& Pandey, S. K., 2005, MNRAS, 358, 968

\bibitem[\protect\citeauthoryear{Bond \etalb}{1991}]{bc91} Bond
J.~R., Cole S., Efstathiou G., Kaiser N., 1991, ApJ, 379, 440

\bibitem[\protect\citeauthoryear{Bowman, Morales, \& Hewitt}{2006}]
{miguel} Bowman J.~D., Morales M.~F., Hewitt J.~N., 2006, ApJ, 638, 20

\bibitem[\protect\citeauthoryear{Ciardi \& Madau}{2003}]
{ciardi} Ciardi B., Madau, P., 2003, ApJ, 596, 1

\bibitem[\protect\citeauthoryear{Furlanetto et al.}{2006}]{fmh06} 
Furlanetto S.~R., McQuinn M., Hernquist L., 2006, MNRAS, 365,
115

\bibitem[\protect\citeauthoryear{Furlanetto et al.}{2006}]{fob06} 
Furlanetto S.~R., Oh S.~P., Briggs, F., 2006, Phys. Rep., 433, 181

\bibitem[\protect\citeauthoryear{Furlanetto et al.}{2004}]{fzh04} 
Furlanetto S.~R., Zaldarriaga M., Hernquist L., 2004, ApJ, 613, 1

\bibitem[\protect\citeauthoryear{Madau \etalb}{1997}]{Madau} 
Madau, P., Meiksin, A., \& Rees, M. J. 1997, ApJ, 475, 429

\bibitem[\protect\citeauthoryear{McQuinn et al.}{2005}]{m05} 
McQuinn M., Furlanetto S.~R., Hernquist L., Zahn O., Zaldarriaga M.,
2005, ApJ, 630, 643

\bibitem[\protect\citeauthoryear{McQuinn et al.}{2006}]{CfA} 
McQuinn M., Zahn O., Zaldarriaga M., Hernquist L., Furlanetto S.~R.,
2006, ApJ, 653, 815

\bibitem[\protect\citeauthoryear{Mellema et al.}{2006}]{mellema}
Mellema G., Iliev I.~T., Pen U.-L., Shapiro P.~R., 2006,
MNRAS, 372, 679

\bibitem[Mo \& White(1996)]{mw96} Mo, H. J. \& White, S. D. M.  1996,
MNRAS, 282, 347

\bibitem[\protect\citeauthoryear{Press \& Schechter}{1974}]{ps74} 
Press W. H., Schechter P., 1974, ApJ, 187, 425

\bibitem[\protect\citeauthoryear{Saiyad-Ali et al.}{2006}]{Ali}
Saiyad-Ali, S., Bharadwaj, S., \& Pandey, S. K., 2006, MNRAS, 366, 213

\bibitem[\protect\citeauthoryear{Scannapieco \& 
Barkana}{2002}]{sb} Scannapieco E., Barkana R., 2002, ApJ, 
571, 585 

\bibitem[Scannapieco \& Thacker(2005)]{st}
Scannapieco E., Thacker R.~J., 2005, ApJ, 619, 1

\bibitem[\protect\citeauthoryear{Sheth \& Tormen}{1999}]{shetht99}
Sheth R.~K., Tormen G., 1999, MNRAS, 308, 119

\bibitem[\protect\citeauthoryear{Sheth \& Tormen}{2002}]{st02} 
Sheth R.~K., Tormen G., 2002, MNRAS, 329, 61

\bibitem[\protect\citeauthoryear{Spergel et al.}{2007}]{Spergel07}
Spergel D.~N., et al., 2007, astro-ph/0603449

\bibitem[\protect\citeauthoryear{Wyithe \& Morales}{2007}]{wyithe}
Wyithe S., Morales M., 2007, astro-ph/0703070

\bibitem[\protect\citeauthoryear{Zahn et al.}{2007}]{zahn}
Zahn O., Lidz A., McQuinn M., Dutta S., Hernquist L., 
Zaldarriaga M., Furlanetto S.~R., 2007, ApJ, 654, 12

\end{thebibliography}
\end{document}